\documentclass[11pt]{article} 

\usepackage{amsmath,amsthm,latexsym,amssymb,amsfonts,epsfig,slashed}

\newtheorem{Theorem}{Theorem}[section]

\newcommand{\be}{\begin{equation}}
\newcommand{\ee}{\end{equation}}
\newcommand{\ba}{\begin{eqnarray}}
\newcommand{\ea}{\end{eqnarray}}

\title{
{\sf 
Hamiltonian Renormalisation VII: Free fermions and doubler free kernels}}
\author{
{\sf T. Thiemann}$^1$\thanks{{\sf 
thomas.thiemann@gravity.fau.de}}\\
\\
{\sf $^1$ Inst. for Quantum Gravity, FAU Erlangen -- N\"urnberg,}\\
{\sf Staudtstr. 7, 91058 Erlangen, Germany}\\
}
\date{{\small\sf \today}}

\makeatletter
\@addtoreset{equation}{section}
\makeatother

\begin{document} 

\maketitle

{\sf

\begin{abstract}
The Hamiltonian renormalisation programme motivated by 
constructive QFT and Osterwalder-Schrader 
reconstruction which was recently launched 
for bosonic field theories is extended to fermions. As fermion quantisation 
is not in terms of measures, the scheme has to be mildly modified accordingly.

We exemplify the scheme for free fermions both for compact and non-compact 
spatial topologies respectively (i.e. with and without IR cut-off) and 
demonstrate that the convenient Dirichlet or Shannon coarse graining kernels 
recently advertised in a companion paper lead to a manifestly doubler free 
flow.   
\end{abstract}

\section{Introduction}
\label{s1}

The Hamiltonian or canonical approach to quantum gravity 
\cite{1} aims at implementing the constraints as operators on a Hilbert
space. In the classical theory, the constraints
generate the Einstein equations via the Hamiltonian equations of motion
\cite{2}. They underlie the numerical implememtation of the initial value 
formulation of Einstein's equations e.g. in black hole merger and 
gravitational wave template codes \cite{3}. 

The mathematically sound construction of canonical quantum gravity is 
a hard problem because the constraints are non-polynomial expressions 
in the elementary fields and in that sense much more non-linear than 
even the most complicated interacting QFT on Minkowski space such as 
QCD whose Hamiltonian is still polynomial in gluon and quark fields.
As the theory is non-renormalisable and thus believed to exist only 
non-perturbatively, the Loop Quantum Gravity (LQG) approach has 
systematically developed such a non-perturbative programme \cite{4}.
LQG derives its name from the fact that it uses a connection rather 
than metric based formulation, hence it is phrased in the language of
Yang-Mills type gauge fields and thus benefits from the non-perturbative 
technology introduced for such theories, specifically gauge invariant 
Wilson loop variables \cite{5}.

The current status of LQG can be described as follows: While the quantum
constraints can indeed be implemented in a Hilbert space representation
\cite{6} 
of the canonical (anti-) commutation and adjointness relations as densely
defined operators \cite{7} and while its commutator algebra is 
mathematically consistent in the sense that it closes, it closes with the 
wrong structure ``functions''. The inverted commas refer to the fact 
that the classical constraints do not form a Lie Poisson algebra because for 
a Lie algebra it is required that one has structure constants. By contrast,
here we have non-trivial structure functions in the classical theory which 
are dictated by the fundamental hypersurface deformation algebra 
\cite{8} and in the quantum theory they become operators
themselves and are not simply constant multiples of the identity operator.
We therefore call them structure operators.

The most important missing step in LQG is therefore to correct those 
structure operators. It is for this reason that more recently Hamiltonian 
renormalisation techniques were considered \cite{9}. There one actually 
works with a 1-parameter family of 
gauge fixed versions of the theory \cite{10} so that the
constraints no longer appear and are traded for a Hamiltonian which drives
that one parameter evolution. The reason for doing is are twofold: On the 
one hand, working with the gauge fixed version means solving the constraints 
classically and saves the work to determine quantum kernel and Hilbert space 
structure on it. On the other hand, the techniques of \cite{9} were derived 
from Osterwalder-Schrader reconstruction \cite{11} which deals with 
theories whose dynamics is driven by an actual Hamiltonian rather than 
constraints (see however \cite{12}).  
Still, that Hamiltonian uniquely descends from 
the constraints and therefore its quantisation implicitly depends on the 
quantisation of the constraints. Therefore, the quantum constrains and 
their structure operators are implicitly also present in the gauge 
fixed version. In addition, in \cite{13} we have shown that the techniques of 
\cite{9} can be ``abused'' also for constrained quantum theories in the sense
that the renormalisation steps to be carried out can be performed 
independently for all constraints ``as if they were actual Hamiltonians'',
even if the corresponding operators are not bounded from below. In that 
sense the methods of \cite{9} complememt those of \cite{14} where the 
correction of the structure operators is approached by exploiting the 
spatial diffeomorphism invariance of the classical theory in an even 
more non-linear fashion than it was alrady done in \cite{7}. 

The programme of \cite{9} rests on the following observation: 
In quantising an interacting classical field theory one cannot proceed
directly but rather has to introduce at least an UV cut-off $M$ where 
we may think of $M^{-1}$ as a spatial resolution. Introducing $M$ produces 
quantisation ambiguities which are encoded in a set of parameters depending 
on $M$.
Almost all points in that set do not define consistent theories where a
consistent theory is defined to be one in which the theory at resolution 
$M$ is the same as the theory at higher resolution $M'>M$ after ``integrating
out'' the extra degrees of freedom. Renormalisation introduces a flow on these 
parameters whose fixed or critical points define consistent theories.
In this way, the correct structure operators or algebra of constraints 
referred to above are also believed to be found, either explicitly or 
implicitly. In \cite{13} we have shown that this is what actually happens 
for the much simpler case of 2d parametrised field theory \cite{13a} whose 
quantum hypersurface deformation algebra coincides with the Virasoro algebra.
The lesson learnt from this is that the quantum constraint algebra {\it 
must not even close} at any finite resolution even if 
{\it the continuum algebra 
closes with the correct structure operators}. In other words, it is 
physically correct that the finite resolution constraints are anomalous 
while the actual continuum theory is anomaly free.\\   
\\     
In \cite{15} we have further tested \cite{9} for free bosons (scalars
and vector fields). Theories with fermions were not considered so far. 
In this paper we close this gap, see also \cite{15a} for a closely 
related formulation.\\
\\
The architecture of this paper is as follows:\\
\\

In section \ref{s2} we briefly recall the bosonic theory from \cite{9}.

In section \ref{s3} we adapt the bosonic theory to the fermionic setting.

In section \ref{s4} we test the fermionic Hamiltonian renormalisation 
theory for free Dirac-Weyl fermions both with and without IR cut-off using 
the Dirichlet-Weyl kernel and confirm a manifestly doubler free spectrum 
at each resolution $M$ at the fixed point. The Nielsen-Ninomiya theorem 
\cite{16} is evaded 
because the finite resolution Hamiltonians are spatially non-local as it 
is usually the case when one ``blocks from the continuum'' i.e. computes 
the ``perfect Hamiltonian''. A similar observation was made in the conext 
of QCD in the Euclidian action approach \cite{17}.

In section \ref{s5} we summarise and conclude.

\section{Review of Hamiltonian renormalisation for bosons}
\label{s2}

To be specific will consider the theory either with IR cut-off so that 
space is a d-torus $T^d$ or without IR cut-off so that space is d-dimensional
Euclidian space $\mathbb{R}^d$ 
and it will be sufficient to consider one coordinate direction as both spaces 
are a Cartesian products. Thus $X=[0,1)$ or $X=\mathbb{R}$ in what follows.

Thus for simplicity we consider a bosonic, scalar 
quantum field $\Phi$ (operator valued distribution) with conjugate 
momentum $\Pi$ on $X$
with canonical commutation and adjointness relations (in natural units 
$\hbar=1$) 
\be \label{3.1}
[\Pi(x),\Phi(y)]=i\; \delta(x,y),\;\;\Phi(x)^\ast=\Phi(x),\;\Pi^\ast(x)=\Pi(x)
\ee
where 
\be \label{3.2}
\delta(x,y)=\sum_{n\in \mathbb{Z}}\; e_n(x)\; e_n(y)^\ast,\; 
e_n(x)=e^{2\pi\;i\;n\;x}
\ee
is the periodic $\delta$ distribution on the torus or 
\be \label{3.2}
\delta(x,y)=\int_{\mathbb{R}}\;\frac{dk}{2\pi} e_k(x)\; e_k(y)^\ast,\; 
e_k(x)=e^{i\;k\;x}
\ee
on the real line respectively. It is customary to 
work with the bounded Weyl operators 
\be \label{3.2}
w[f,g]=\exp(i[\Phi(f)+\Pi(g)]),\;\;
\Phi(f)=\int_{X}\; dx\; f(x)\; \Phi(x),\;
\Pi(g)=\int_{X}\; dx\; g(x)\; \Pi(x)
\ee
with $f,g\in L=L_2(X,dx)$ test functions or smearing functions 
usually with some additional properties such 
as differentiability or even smoothness. For tensor fields of higher degree
a similar procedure can be followed (see \cite{9}).

Since the space $L$ enters the stage naturally we use multi resolution 
analysis (MRA) language \cite{18} familiar from wavelet theory \cite{19}
to define a renormalisation group flow. MRA's serve as a powerful organising
principle to define renormalisation flows in terms of coarse graining 
kernels and while the choice of the kernel should intuitively 
not have much influence
on the fixed point or continuum theory (at least in presence of universality)
the examples of \cite{13,18b} show that generic features such as smoothness
can have an impact.  

In the most general sense 
an MRA is a nested sequence of Hilbert subspaces $V_M\subset L$ indexed
by $M\in {\cal M}$ where $\cal M$ is partially ordered and directed 
by $\le$. That is, one has $V_M\subset V_{M'}$ for $M\le M'$ and 
$\cup_{M\in {\cal M}}\; V_M$ is dense in $L$. Pick an ONB 
$d(M)^{1/2}\;\chi^M_m$ for $V_M$ where $m$ is from a countably 
finite (infinite) index set $Z_M$ for $X=[0,1)$ ($X=\mathbb{R}$) respectively
and $d(M)$ is a finite number. In case that $X=[0,1)$ typically $Z_M$ is 
the lattice $x^M_m,\; m/d(M)$ and $d(M)=\dim(V_M)$ the number of points in 
it.
Let $L_M=l_2(Z_M)$ be the Hilbert space of square summable sequences indexed 
by $Z_M$ with inner product 
\be \label{3.3}
<f_M,g_M>:=d(M)^{-1}\;\sum_{m\in Z_M} \; f^\ast_M(m)\; g_M(m)
\ee
This scalar product offers the interpretation of $f_M(m):=f(x^M_m),\; 
x^M_m:=\frac{m}{d(M)}$ and similar for $g_M$
as the discretised values of some 
functions $f,g\in L$ in which case (\ref{3.3}) is the Riemann sum 
approximant of $<f,g>_L$. It is for this reason that we did not normalise
the $\chi^M_m$. 

What follows works for any such choice of ONB indexed by $M$. 
However, to reduce the amount of arbitrariness and to give additional 
structure to MRA's one requires, both in wavelet theory and renormalisation,
in addition that the ONB's descend from a few mother scaling functions 
$\phi$ by dilatations depending on $M$ and translations depending on $m$. In 
wavelet theory on the real line 
one is rather specific about the concrete desecendance. 
In particular, there is only one mother scaling 
function, the $\chi^M_m$ and $\phi$ are linearly related, $\cal M$ just 
consists of the powers $M=2^N,\; N\in \mathbb{Z}$ and 
$\chi^M_m=\phi(M\;x-m)$. As 
advertised in \cite{18} we allow a more general descendance and thus 
accept a finite, fixed number of mother scaling functions and that 
the $\chi^M_m$ are dilatations and translations of a rational function 
of those mother scaling functions. This keeps the central idea of providing 
minimal structure to an MRA while increasing flexibility.

The spaces $V_M, L_M$ are in bijection via
\be \label{3.3a}
I_M:\; L_M\to L,\;f_M\mapsto \sum_m\; f_M(m)\; \chi^M_m
\ee
Note that (\ref{3.3a}) has range in $V_M\subset L$ only. 
Its adjoint $I_M^\dagger:\; L\to L_M$ is defined by 
\be \label{3.4}
<I_M^\dagger f,f_M>_{L_M}:=<f,\; I_M\; f_M>_{L}
\ee
so that 
\be \label{3.5}
(I_M^\dagger f)(m)=d(M)\; <\chi_M,f>_L
\ee
Clearly 
\be \label{3.6}
(I_M^\dagger I_M f_M)(m)=d(M)\; <\chi^M_m, I_M f_M>_L =f_M(m)
\ee 
i.e. $I_M^\dagger I_M=1_{L_M}$ while 
\be \label{3.7}
(I_M I_M^\dagger f)(x)=d(M)\; \sum_m\; \chi^M_m(x) <\chi^M_m,f_M>_L
=(p_M f)(x)
\ee
is the projection $P_M:\; L\mapsto V_M$.

Given $M\le M'$ we define the coarse graining map
\be \label{3.8}
I_{MM'}:=I_{M'}^\dagger \; I_M:\; L_M\mapsto L_{M'}
\ee
It obeys 
\be \label{3.9}
I_{M'}\; I_{MM'}=p_{M'}\; I_M=I_M
\ee  
because $I_M$ has range in $V_M\subset V_{M'}$ for $M\le M'$. 
This is the place where the MRA property of the nested set of subspaces 
$V_M$ was important. Next for $M_1\le M_2\le M_3$ we have 
\be \label{3.9}
I_{M_2 M_3}\; I_{M_1 M_2}=
I_{M_3}^\dagger\; p_{M_2}\; I_{M_1}=  
I_{M_3}^\dagger\; I_{M_1}=I_{M_1 M_3}
\ee
for the same reason. This is called the condition of cylindrical 
consistency which is crucial for the renormalisation group flow.

To see the importance of (\ref{3.9}) we consider a probability 
measure $\nu$ on the 
space $\cal F$ of field configurations $\Phi$ which defines a Hilbert space   
${\cal H}=L_2({\cal F},d\nu)$ and a representations space for the Weyl
algebra $\mathfrak{A}$ generated from  the Weyl elements (\ref{3.2}). We 
set $w[f]:=w[f,g=0]$ and define the generating functional of moments of $\nu$
by 
\be \label{3.10}
\nu(f):=\nu(w[f])
\ee
If we restrict $f$ to $V_M$ we obtain an effective measure on the space 
of discretised quantum fields $\Phi_M=I_M^\dagger \Phi$ via 
\be \label{3.11}
w[I_M f_M]=w_M[f_M]=e^{i\Phi_M(f_M)},\; \Phi_M(f_M)=<f_M,\Phi_M>_{L_M}
\ee
and 
\be \label{3.12}
\nu_M(f_M):=\nu(w[I_M f_M])=\nu_M(w_m[f_M])
\ee
The measures $\nu_M$ on the spaces ${\cal F}_M$ of fields $\Phi_M$ are 
consistently defined by construction
\be \label{3.13}
\nu_{M'}(I_{MM'} f_M)=\nu_M(f_M)
\ee
for any $M<M'$ since the $\nu_M$ descend from a continuum measure.
Conversely, given a family of measures $\nu_M$ satisfying
(\ref{3.13}) a continuum measure $\nu$ can be constructed known as the 
projective limit of the $\nu_M$ under mild technical assumptions 
\cite{20}. 
To see the imprortance of (\ref{3.9}) for this to be the case,
suppose we write $f\in L$ in two eqivalent ways 
$f=I_{M_1} f_M=I_{M_2} g_{M_2}$ then we should have 
$\nu_{M_1}(f_{M_1})=\nu_{M_2}(g_{M_2})$. Now while $M_1,M_2$ may not be 
in relation, as $\cal M$ is directed we find $M_1,M_2\le M_3$. Applying 
$I_{M_3}^\dagger$ we conclude $I_{M_1 M_3} f_{M_1}=I_{M_2 M_3} g_{M_2}$ thus 
due to (\ref{3.13}) indeed
\be \label{3.14}
\nu_{M_1}(f_1)=\nu_{M_3}(I_{M_1 M_3} f_{M_1})=
\nu_{M_3}(I_{M_2 M_3} g_{M_2})=
\nu_{M_2}(g_{M_2})
\ee
In CQFT the task is to construct a representation of the Weyl algebra
$\mathfrak{A}$ with 
additional properties such as allowing for the imlementation of a 
Hamiltonian operator $H=H[\Phi,\Pi]$ which imposes severe restrictions 
on the Hilbert space representation. One may start with discretised 
Hamiltonians 
\be \label{3.15}
H^{(0)}_M:[\Phi_M,\Pi_M]:=H(p_M\Phi,p_M\Pi]
\ee
on ${\cal H}^{(0)}_M:=L_2({\cal F}_M,\nu^{(0)}_M)$ where $\nu^{(0)}_M$ is 
any probability measure to begin with, for instance a Gaussian measure
or a measure constructed from the ground state $\Omega^{(0)}_M$ of the 
Hamiltonian $H^{(0)}_M$. The point of using the 
IR cut-off is that there are only finitely 
many, namely $d(M)$ degrees of freedom $\Phi_M,\Pi_M$ which are conjugate 
\be \label{3.16}
[\Pi_M(m),\Phi(m')]=i\; d(M)\; \delta(m,m'),\;\;\Phi_M(m)^\ast=
\Phi_M(m),\;\Pi_M^\ast(m)=\Pi_M(m)
\ee
so that construction of $\nu^{(0)}_M$ does not pose any problems. 
In case that there is no IR cut-off it is significantly harder to 
show that the theories even at finite UV cut-off exist. Assuming this 
to be the case, one 
fixes for each $M\in {\cal M}$ an element $M\le M'(M)\in {\cal M}$ and 
defines isometric injections 
\be \label{3.17}
J^{(n+1)}_{MM'(M)}:\; {\cal H}^{(n+1)}_M\to {\cal H}^{(n)}_{M'(M)},\;\;
{\cal H}^{(n)}_M:=L_2({\cal F}_M,d\nu^{(n)}_M)
\ee
via 
\be \label{3.18}
\nu^{(n+1)}_M(f_M):=\nu^{(n)}_{M'(M)}(I_{MM'(M)} f_M)
\ee
and with these the flow of Hamiltonians
\be \label{3.19}
H^{(n+1)}_M:=J_{MM'(M)}^\dagger\; H^{(n)}_{M'(M)}\; J_{MM'(M)}
\ee
The isometry of the injections relies on the assumption that the span of the 
$w_M[f_M]$ is dense in ${\cal H}^{(0)}_M$ which is typically the case. 

This defines a sequence or flow (indexed by $n$) of families (indexed by $M$) 
of theories ${\cal H}^{(n)}_M,H^{(n)}_M$. At a critical or fixed point of 
this flow the consistency condition (\ref{3.13}) is satisfied (at first 
in the linearly ordered sets of ${\cal M}(M):=\{(M')^N(M),\; N\in
\mathbb{N}_0\}$ and then ususally for all of $\cal M$ by universality) 
and one obtains a consistent family $({\cal H}_M,\;H_M)$. This family
defines a continuum theory $({\cal H}, H)$ as one obtains 
inductive limit 
isometric injections $J_M:\; {\cal H}_M \mapsto {\cal H}$ such that 
$J_{M'} J_{MM'}=J_M,\;M\le M'$ thanks to the fixed point identiy  
$J_{M_2 M_3}\;J_{M_1 M_2}=J_{M_1 M_3},\;M_1\le\; M_2\le M_3$ and such that
\be \label{3.20}
H_M=J_M^\dagger \; H\; J_M
\ee
is a consistent family of quadratic forms 
$H_M=J_{MM'}^\dagger\; H_{M'}\; J_{MM'},\; M\le M'$.\\
\\
We conclude this section by noting that wavelet theory actually also 
seeks to decompose the spaces as $V_{M'}=V_M\oplus W_M$ where $W_M$ is 
the orthogonal complement of $V_M$ in $V_{M'},\;M\le M'$, to provide an 
ONB for the $W_M$ and to require that this basis descends from a mother 
wavelet $\psi$ concretely related to the scaling function 
in the same specific way as outlined above for the scaling function. For
the purpose of renormalisation this additional structure is not essential,
thus we will not go int further details. We remark however that in \cite{18}
we also generalised the notion of wavelets in the same way as for the scaling
function which again keeps the central idea of structurising the MRA and 
showed that the Dirichlet and Shannon kernels are non-trivial realisations 
of that more general definition.

\section{Hamiltonian renormalisation for Fermions}
\label{s3}

To distinguish the bosonic field $\Phi$ from the previous section 
form the present fermionic field we use the notation $\xi_B$ for a chiral 
(or Weyl) 
fermion where $B=1,2$ transforming in one of the two fundamental 
representations of $SL(2,\mathbb{C})$. Its Majorana conjugate 
$\epsilon\xi^\ast$ with $\epsilon=i\sigma_2$ (Pauli matrix) 
of opposite chirality then transforms 
in the dual fundamental representation. We have the fundamental simultaneous
canonical anti-commutation relations (CAR)
\be \label{6.31a}
[\xi_B(x),\xi_C(y)^\ast]_+ :=
\xi_B(x)\;\xi_C(y)^\ast + \xi_C(y)^\ast \; \xi_B(x)
=\delta_{BC}\; \delta(x,y)
\ee
all other anti-commutators vanishing. Dirac fermions and Majorana fermions 
can be considered as usual by using direct sum $SL(2,\mathbb{C})$ 
representations of 
 of independent Weyl fermions 
of opposite chirality or of the direct sum of a fermion with its Majorana 
conjugate. It will be sufficient to consider a single Weyl fermion species 
$\xi_B$ for what follows.   

The measure theoretic 
language given for bosons of the previous section
cannot apply because of several reasons: i. the ``Weyl elements''
$w[f]:=\exp(i\;<f,\xi>_L^\dagger)$ are not mutually commuting, ii. the 
$w[f]\Omega$ are not dense in the Fock space $\cal H$ 
defined by $<f,\xi>\Omega=0$
because in fact $w[f]=1_{{\cal H}}+i<f,\xi>^\dagger$ due to nil-potency and 
iii. $w[f]$ is not unitary.
To avoid this one can formally work with Berezin ``integrals'' \cite{21} and 
anti-commuting smearing fields $f$ but then we cannot immediately transfer 
the functional analytic properties of the commuting test functions from 
the bosonic theory 
and apart from serving as compact organising tool, anticommuting smearing 
functions do not 
have any advantage over what we say below.

One of the motivations to work with Weyl elements rather than say 
$\Phi(f),\Pi(f)$ in the bosonic case
is that the Weyl elements are bounded operators.
However, the operators $\xi(f):=<f,\xi>_L,\;\xi(f)^\dagger$ are 
already bounded by $||f||_L$ as follows from the CAR
\be \label{6.31}
[\xi(f),\xi(f)^\ast]_+=||f||_L^2\; 1_{{\cal H}}\;\;\Rightarrow\;\;   
||\xi(f)\;\psi||_{{\cal H}}^2,\;||\xi(f)\;\psi||_{{\cal H}}^2
\; \le\; ||f||_L^2\; ||\psi||^2
\ee
The derivation of the renormalisation scheme given in \cite{9} in fact covers 
both the bosonic and fermionic case but the practical implememtation for 
bosons used measures \cite{15}.
We thus adapt the bosonic renormalisation scheme by reformulating it in
an eqivalent way which then extends to the fermionic case:\\
Given cyclic vectors $\Omega^{(n)}_M$ for the algebra generated by 
the 
\be \label{6.31a}
\xi_M(f_M):=<I_M f_M,\xi>_{L}=<f_M,\;I_M^\dagger \xi>_{L_M}=
\frac{1}{d(M)}\;\sum_{B,m\in Z_M}\; [f^B_M(m)]^\ast\; \xi_{M,B}(m)
\ee
and their adjoints (perhaps the vacua of the Hamiltonians $H^{(n)}_M$) 
we define the flow of isometric injections (e.g. for $M'=M'(M)$)
\be \label{6.32}
J^{(n+1)}_{MM'}\;\Omega^{(n+1)}_M:=\Omega^{(n)}_{M'},\;\;
J^{(n+1)}_{MM'}\;
\Xi_M(F_{M,1})..\Xi_M(F_{M,N})\;\Omega^{(n+1)}_M
:=
\Xi_{M'}(I_{MM'} F_{M,1})^\ast..
\Xi_{M'}(I_{MM'} F_{M,N})^\ast\;\Omega^{(n)}_{M'}
\ee
Note that $\xi_M=d(M)\; I_M^\dagger\xi$ preserve the CAR in the sense that 
\be \label{6.32a}
[\xi_M(m),\;[\xi_M(m')]^\ast]_+=d(M)\; M\delta_{mm'}
\ee
and 
$\Xi(F)=\sum_B\; [<f_B,\xi_B>_L + <\tilde{f}_B,\xi_B>^\ast]$ where we have 
collected four independent smearing functions $f_B,\tilde{f}_B,\;B=1,2$ into 
one symbol $F$. The same notation was used in (\ref{6.32}) for the $M$ 
dependent quantities. With these we define the flow of Hamiltonian 
quadratic forms as
\be \label{6.33}  
H^{(n+1)}_M:=
[J^{(n+1)}_{MM'}]^\dagger\; H^{(n)}_{M'}\; J^{(n+1)}_{MM'}
\ee
These formulas are even simpler than in the bosonic case because there is 
no fermionic Gaussian measure and corresponding covariance to consider. 
However, as in the bosonic case, one has to give initial data for this 
flow. This can be done, e.g. by defining 
\be \label{6.33a}
H^{(0)}_M[\xi_M,\xi_M^\ast]:=
\;:\; H[p_M \xi,\;(p_M\xi)^\ast] \; : \;
\ee
where $(p_M \xi)_B:=I_M\;I_M^\dagger \xi_B$, $H$ is the classical Hamiltonian 
and $:.:$ denotes normal ordering with respect to a Fock space 
${\cal H}^{(0)}_M$ with cyclic Fock vacuum $\Omega^{(0)}_M$ annihilated 
by $A^{(0)}_{B,M}$ assembelled from $\xi_{M,B},\;\xi_{M,B}^\ast$ as suggested
by the form of $H[p_M \xi,\;(p_M\xi)^\ast]$. As in the bosonic case, the 
fields $\xi_{M,B}$ do not depend on the sequence label $n$ while the 
annihilators $A^{(n)}_{M,B}$ do as one obtains them from the $\xi_{M,B}$ 
using extra discretised structure that depends on $M$, typically 
lattice derivatives
and more complicated aggregates made from those (Dirac-Weyl operators, 
Laplacians,..).

\section{Hamiltonian renormalisation of free fermions and fermion doubling}
\label{s4}

In this section we will concretely choose the renormalisation structure as 
follows (see \cite{18} for more details): $Z_M$ will be the lattice of 
points $x^M_m$ with $m\in \mathbb{Z}$
if $X=\mathbb{R}$ and $m\in \mathbb{Z}_M:=\{0,1,2,..,M-1\}$ if 
$X=[0,1)$ respectively and $d(M)=M$. The set $\cal M$ consists 
of the odd naturals with partial order $M\le M'$ iff $M'/M\in \mathbb{N}$.
The renormalisation sequence will be constructed using $M'(M)=3M$ for 
simplicity. The MRA's are based on the Shannon \cite{21} 
and \cite{22} Dirichlet kernels 
respectively, that is,
\be \label{6.0}
\chi^M_m(x)=
\left\{ \begin{array}{cc} 
\frac{\sin(M\;\pi\;(x-x^M_m))}{M\;\pi\; (x-x^M_m)} & X=\mathbb{R} \\
\frac{\sin(M\;\pi\;(x-x^M_m))}{M\;\sin(\pi (x-x^M_m))} & X=[0,1) 
\end{array}
\right.
\ee
Their span is dense in $V_M$ and they are mutually orthogonal with norm 
$M^{-1}$. The Dirichlet kernel is 1-periodic as it should be. Both have 
maximal value 1 at $x=x^M_m$, are symmetric about this point and (slowly)
decay away from it, thus display some position space locality. They are 
real valued and smooth and have compact momentum support $k\in 
[-\pi M,\pi M]$ and $k=2\pi n,\; n\in \hat{\mathbb{Z}}_M=\{-\frac{M-1}{2},
-\frac{M-1}{2}+1,..,\frac{M-1}{2}\}$ respectively.

Recall the following facts about the topologies of position space and momentum
space via the Fourier transform where we denote by $M$ the spatial resolution
of the lattice $x^M_m$ with either $m\in \mathbb{Z}$ or 
$m\in\mathbb{Z}_M=\{0,1,2,..,M-1\}$ where for $M$ odd we set  
$\hat{\mathbb{Z}}_M=\{-\frac{M-1}{2},..,\frac{M-1}{2}\}$ 
(c: compact, nc: non-compact, d: discrete, nd: non-discrete (continuous)):
\be \label{6.1}
\begin{array}{ccc}
{\sf space-topology} & {\sf momentum-topology} & {\sf Fourier-function}\\
& \\
{\sf nc,~~ nd:}\;\;\mathbb{R} 
&  {\sf nc,~~ nd:}\;\;\mathbb{R}
& e_k(x)=e^{i\;k\;x}
\\
{\sf nc,~~ d:}\;\;\frac{1}{M}\cdot\mathbb{Z} 
&  {\sf c,~~ nd:}\;\;[-M\pi,\;M\pi) 
& e^M_k(m)=e^{i\;k\; x^M_m}
\\
{\sf c,~~ nd:}\;\;[0,1) 
&  {\sf nc,~~ d:}\;\;\mathbb{Z} 
& e_n(x)=e^{2\pi\; i\; n\;x}
\\
{\sf c,~~ d:}\;\;\frac{1}{M}\cdot \mathbb{Z}_M 
&  {\sf c,~~ d:}\;\;\;\hat{\mathbb{Z}}_M
& e^M_n(m)=e^{2\;\pi\;i\;n\;x^M_m} 
\end{array}
\ee
Accordingly, in the non-compact and comact case respectively, 
the space of Schwartz test functions 
is a suitable subspace of $L=L_2(\mathbb{R},dx)$ and $L=L_2([0,1),dx)$
respectively which have momentum support in $2\pi\mathbb{R}$ and
$2\pi\cdot \mathbb{Z}$ respectively. Upon discretising space into cells 
of width $1/M$ the momentum support $\mathbb{R}$ and $\mathbb{Z}$ respectively 
gets confined to the Brillouin zones 
$[-\pi\;M,\pi M)$ and $\hat{\mathbb{Z}}_M$ respectively.

The corresponding completeness relations or resolutions of the identity 
read
\ba \label{6.2}
\delta_{\mathbb{R}}(x,x') &=& \int_{\mathbb{R}}\;\frac{dk}{2\pi}\; e_k(x-x')
\nonumber\\
M\;\delta_{\mathbb{Z}}(m,m') 
&=& \int_{-\pi \;M}^{\pi M}\; \frac{dk}{2\pi}\; e^M_k(m-m')
\nonumber\\
\delta_{[0,1)}(x,x') &=& 
\sum_{n\in \mathbb{Z}}\; e_n(x-x')
\nonumber\\
M\;\delta_{\mathbb{Z}_M}(m,m') &=& \sum_{n\in \mathbb{Z}_M}\; e^M_n(m-m')
\ea
While the first and third relation in (\ref{6.2}) define the $\delta$ 
distribution on $\mathbb{R}$ and $[0,1)$ respectively, the second and fourth
relation in (\ref{6.2}) are the restrictions to the lattice of the regular
functions 
\ba \label{6.3}
\delta_{\mathbb{R},M}(x) &=& \int_{-\pi \;M}^{\pi M}\; 
\frac{dk}{2\pi}\; e_k(x)=\frac{\sin(\pi\;M\;x)}{\pi\; x}
\nonumber\\
\delta_{[0,1),M}(x) &=& \sum_{n\in \mathbb{Z}_M}\; e_n(x) 
=\frac{\sin(\pi\;M\;x)}{\sin(\pi\; x)}
\ea
which we recognise as the Shannon (sinc) and Dirichlet kernel respectively.
After dividing and dilating them by $M$ and tranlating them by $m$
we obtain precisely the functions (\ref{6.0}). These 
kernels can be considered as regularisations of the afore mentioned 
$\delta$ distributions in the sense that the momentum integral 
$k\in \mathbb{R}$ or momentum sum $n\in \mathbb{Z}$ has been confined to
$|k|<\pi M$ and $|n|<\frac{M-1}{2}$ respectively. Both are real valued, 
smooth, strongly peaked at $x=0$ and have compact momentum support. 
The Shannon kernel like the Dirichlet kernel 
is an $L_2$ function but it is not of rapid decay with respecto to 
position.\\   
\\
The simplest possible action for fermions is the massless,
chiral theory in 2d Minkowski space 
\be \label{6.11}
S=i\int_{\mathbb{R}}\; dt\; \int_X\; dx\; \bar{\xi}\slashed{\partial}\xi
\ee
Here $X=\mathbb{R}$ or $X=[0,1)$. The 2d Clifford algebra with signature 
$(-1,+1)$ is generated by $\gamma^0=\epsilon=i\sigma_2,\; 
\gamma^1=\sigma_1$ where 
$\sigma_1, \sigma_2,\sigma_3=\epsilon\sigma_1$ are the Pauli matrices. Then 
$\slashed{\partial}=\gamma^\mu\partial_\mu,\; x^0=t, x^1=x$ and 
$\bar{\xi}=(\xi^\ast)^T\; \gamma^0$. Due to 
$([\gamma^0 \gamma^\mu]^\ast)^T=\gamma^0\gamma^\mu$ the action is real valued.
Generalisations to higher dimensions, massive theories, with more species 
or higher spin are immediate and just require the corresponding Clifford 
algebras.  

Then $i[\xi^A]^\ast,\; A=1,2$ is canonically 
conjugate to $\xi^A$ which results in the 
non vanishing canonical anti commutation relations (CAR)
\be \label{6.12}
[\xi^A(x),(\xi^B)^\ast(y)]_+=\delta^{AB}\;\delta(x,y)
\ee
and the Hamiltonian is 
\be \label{6.13}
H=-i\;\int_X\;dx\; \{[\xi^\ast]^T\sigma_3\; \xi'\}(x)
\ee
with $\xi'=\partial\xi/\partial x$ which is linear in spatial derivatives.
Indeed the Dirac-Weyl equation $\slashed{\partial}\xi=0$ is reproduced by 
the Heisenberg equation of (\ref{6.13})
\be \label{6.13a}
i\dot{\xi}=[H,\xi]=i\sigma_3 \xi'\;\;\Leftrightarrow\;\;
\epsilon\dot{\xi}-\epsilon\sigma_3\xi'=\slashed{\partial}\xi=0
\ee
As (\ref{6.13}) is indefinite as it stands we introduce the self-adjoint 
projections on $L=L_2(X,dx)$ with $s=\pm 1$ 
\be \label{6.14}
Q_s=\frac{1}{2}[1_L+i\;s\frac{\partial}{\omega}]\;Q,\;
Q=1_L-1\;<1,.>_L/||1||_L^2,\;\omega=\sqrt{-\partial^2},\;\;
i\partial Q_s=s\; \omega Q_s
\ee
Note $Q=1_L$ for $X=\mathbb{R}$. We then rewrite the Hamiltonian as 
\ba \label{6.15}
-H &=& 
<\xi_1,[Q_+ -\;Q_-]\omega\;\xi_1>_L
-<\xi_2,[Q_+ -\;Q_-]\omega\;\xi_2>_L
\\
&=&
<Q_+ \xi_1,\;\omega\;Q_+\xi_1>_L
-<Q_- \xi_1,\;\omega\;Q_-\xi_1>_L
-<Q_+ \xi_2,\;\omega\;Q_+\xi_2>_L
+<Q_- \xi_2,\;\omega\;Q_-\xi_2>_L
\nonumber
\ea
Thus we declare 
\be \label{6.14a}
A_{1,+}:=(Q_+ \xi_1)^\ast,\; A_{1,-}:=Q_- \xi_1,
A_{2,-}:=(Q_- \xi_2)^\ast,\; A_{2,+}:=Q_+ \xi_2
\ee 
as annihilators 
and obtain the normal ordered, positive semi-definite Hamiltonian 
\be \label{6.16}
:H:=\sum_{B=1,2;\sigma=\pm}\; \int_X\; dx\; A_{B,\sigma}^\ast \;\omega\;
A_{B,\sigma}
\ee
where the $A_{B,\sigma}$ obey the CAR
\be \label{6.17}
[A_{B,\sigma}(x),[A_{B',\sigma'}(x')]^\ast]_+=
\delta_{BB'}\;\delta_{ss'}\;Q_s(x,x')
\ee
where $Q_s(x,x')$ is the integral kernel $(Q_s f)(x)=\int_X\; dx'\; 
Q_s(x,x') \; f(x')$. Note that the zero modes of $\xi_B$ do no not contribute 
to $H$ so we have to quantise them without guidance from the form of 
the Hamiltonian. With $Q^\perp=1_L-Q$ we 
define $A_{B,0}:=Q^\perp \xi_B$ as the annihilation operator which is non 
vanishing only for $X=[0,1)$. 
 
From this perspective, 
the problem of the fermion doublers on the lattice $\frac{1}{M}\mathbb{Z}$ or 
$\frac{1}{M}\mathbb{Z}_M$ for $X=\mathbb{R}$ and $X=[0,1)$ respectively
is encoded in the way one 
discretises the partial derivative $\partial$ that appears in the 
projections $Q_s$ (in Hamiltonian renormalisation the time variable 
and time derivatives are kept continuous). 
For scalar theories, $\partial$ appears only quadratically
in the Laplacian $\Delta=-\partial^2$ while for fermions it appears linearly.
This problem is therefore not only present for fermions but for all theories 
in which besides the Laplacian also the partial derivatives themselves 
are involved in the quantisation process. One such example is parametrised 
field theory which shares many features with string theory \cite{11}.

Alternatively, this problem shows up in the discretisation of the 2-point 
functions of the theory (as the theory is free, the two point function 
determines all higher N-point functions). To compute them from the current
Hamiltonian setting we use the CAR to compute the Heisenberg
time evolution of the 
annhilators (from now on normal ordering is being understood) 
\be \label{6.17a}
A_{B,\sigma}(t,x)=e^{-it H}\; A_{B,\sigma}(x)\; e^{it H}=
[e^{it \omega} A_{B,\sigma}](x)
\ee
where $Q_\sigma A_{B,\sigma}=A_{B,\sigma}$ was used. Then the non-vanishing 
two point functions are 
\ba \label{6.17b}
&& <\Omega, \xi_B(s,x)\; \xi_C(t,y)^\ast\;\Omega>
=<\Omega, ([Q_+ + Q_- + Q^\perp]\xi_B)(s,x)\; 
([Q_+ + Q_- + Q^\perp]\xi_C](t,y)^\ast\;\Omega>
\nonumber\\
&=& <\Omega, 
\{\delta_{B,1}[A_{1,+}^\ast + A_{1,-} + A_{1,0}]
+\delta_{B,2}[A_{2,+} + A_{2,-}^\ast + A{2,0}]\}(s,x)\;\times
\nonumber\\
&& \{\delta_{C,1}[A_{1,+} + A_{1,-}^\ast + A_{1,0}^\ast]
+\delta_{C,2}[A_{2,+}^\ast + A_{2,-}+A_{2,0}^\ast]\}(t,y)\;\Omega>
\nonumber\\
&=& <\Omega, 
\{\delta_{B,1}\; [A_{1,-}+A_{1,0}] + \delta_{B,2} [A_{2,+}+A_{2,0}]\}(s,x)\;
\{\delta_{C,1}\;[A_{1,-}^\ast+A_{1,0}^\ast] +
\delta_{C,2}\; [A_{2,+}^\ast + A_{2,0}^\ast\}(t,y)\;\Omega>
\nonumber\\
&=& e^{is\omega_x-it \omega_y}\;
\;
\{
\delta_{1,B}\delta_{1,C}\;[Q_-(x,y)+Q^\perp](x,y)
+\delta_{2,B}\delta_{2,C}\;Q_+(x,y)+Q^\perp](x,y)
\}
\nonumber\\
&=& \frac{1}{2} \;e^{is\omega_x-it \omega_y}\;
\;
\{\delta_{BC}(1+Q^\perp)-i\;[\sigma_3]_{BC}\frac{\partial_x}{\omega_x}\}
\delta(x,y)
\nonumber\\
&=& \frac{\delta_{BC}}{2\;||1||^2}
+\int\; \frac{dk}{2\pi\;2\omega(k)}\; e^{i[\omega(k)(s-t)-k(x-y)]}
[\omega(k)\; 1_2-k\sigma_3]_{BC}
\nonumber\\
&=& 
\frac{\delta_{BC}}{2\;||1||^2}
+\int\; \frac{dk}{2\pi\;2\omega(k)}\; e^{-i\; K\cdot(X-Y)}
[K^0\;(1+Q^\perp) 1_2-K^1\sigma_3]_{BC}
\nonumber\\
&=& \frac{\delta_{BC}}{2\;||1||^2}
-i\;
[1_2(1+Q^\perp) \partial_{X^0}+\sigma_3\;\partial_{X^1}]_{BC}
\;\int\; \frac{dk}{2\pi\;2\omega(k)}\; e^{-i\; K\cdot(X-Y)}
\nonumber\\
&=& \frac{\delta_{BC}}{2\;||1||^2}
+ i\;([\epsilon
(1+Q^\perp)\partial_{X^0}+\sigma_1\;\partial_{X^1}]\;\epsilon)_{BC}
\Delta_+(x-y)
\nonumber\\
&=& \frac{\delta_{BC}}{2\;||1||^2}
+i\;
[\slashed{\partial}_X \; \epsilon]_{BC}
\Delta_+(X-Y)
\ea
with $K^0:=\omega(k)=|k|,\; K^1=k$ and $X^0=s,\; X^1=x, \;Y^0=t,\; Y^1=y$
and $K\cdot X=-K^0 X^0 +K^1 X^1$. Here $\Delta_+$ is the Wightman two 
point function of a free massless Klein-Gordon field in 2d Minkowski space
\be \label{6.17c}
\Delta_+(X-Y)=
\int\; \frac{dk}{2\pi\;2\omega(k)}\; e^{-i\; K\cdot(X-Y)}
\ee
A similar computation yields ($X,Y$ and $B,C$ and $Q_+, Q_-$ switch and 
the contribution from $A_{B,0}$ is missing leading to $-\delta_{BC}$ in the 
final result) 
\be \label{6.17d}
<\Omega, \; \xi_C(t,y)^\ast\;\xi_B(s,x)\;\Omega>
=-\frac{\delta_{BC}}{2\;||1||^2}
+i\epsilon[\epsilon \partial_{Y^0}-\sigma_3 \partial_{Y^1}]_{CB}
\;\Delta_+(Y-X)
=-\frac{\delta_{BC}}{2\;||1||^2}
+i[\epsilon\; \slashed{\partial}_Y]_{CB}\; \Delta_+(Y-X)
\ee
Using the conjugate spinor $\overline{\xi}=[\xi^\ast]^T\epsilon$ we may 
rewrite (\ref{6.17c}), (\ref{6.17d}) as 
\ba \label{6.17e}
<\Omega,\;\xi(X)\otimes \overline{\xi}(Y)\;\Omega>x=
&=& \frac{\epsilon}{2\;||1||^2}
+i\;\slashed{\partial}_X\; \Delta_+(X-Y),\;\;       
\nonumber\\
<\Omega,\;\overline{\xi}(Y)\otimes \xi(X)\;\Omega>x
&=& -\frac{\epsilon}{2\;||1||^2}
+i\;\slashed{\partial}_Y\; \Delta_+(Y-X),\;\;       
\ea
which gives the time ordered 2 point function or Feynman propagator 
\ba \label{6.17f}
&& D_F(X-Y):=<\Omega,\;T[\xi(X)\otimes \overline{\xi}(Y)]\;\Omega>
\nonumber\\
& :=&
\theta(X^0-Y^0)\;
<\Omega,\;\xi(X)\otimes \overline{\xi}(Y)\;\Omega>
-\theta(Y^0-X^0)\;
<\Omega,\;\overline{\xi}(Y)\otimes \xi(X)\;\Omega>
\nonumber\\
&=&\slashed{\partial}_X \Delta_F(X-Y)
\ea
%
%
%
where 
\be \label{6.17g}
\Delta_F(X-Y)=-i\;\lim_{\epsilon\to 0+}\;\int\; \frac{d^2K}{(2\pi)^2}\; 
\frac{e^{-iK\cdot(X-Y)}}{-K\cdot K-i\epsilon}
\ee
is the Feynman propagator of the 2d massless Klein Gordon field. We see 
that $\slashed{\partial}_X \;D_F(X-Y)=i\delta^{(2)}(X-Y)$ due to 
$\slashed{\partial}^2=\Box$, i.e. $D_F=i\slashed{\partial}^{-1}$. In 
Hamiltonian renormalisation one discretises only $x,\partial_x$ and confines 
only $|K^1|<\pi M$ while in the Euclidian approach one discretises also
$t,\partial_t$ and confines $|K^0|<\pi M$. In any case we see that it is the 
projections $Q_s$ that directly translate into $\slashed{\partial}$ which 
is linear in the derivatives. If the propagator is to keep the property 
to invert the Dirac-Weyl operator $\slashed{\partial}$ then we are forced 
to write the momentum expression of (\ref{6.17f}), say in the Hamiltonian
approach, as  
\be \label{6.17h}
\frac{\epsilon\; K_0+\sigma_1\;\lambda_M(K_1)}
{K_0^2-\lambda_M(K_1)^2-i\epsilon}
\ee
where $[\partial_M e_{K_1}](X^1)=i\lambda_M(K_1)\;e_{K_1}(X^1),\; 
X^1\in\mathbb{Z}/M$ defines the 
eigenvalues of the discrete derivative and indices are moved with 
the Minkowski metric. 

The case $X=[0,1)$ is literally the same, just that we must sum over 
$k=K^1=2\pi n,\;n\in \mathbb{Z}$ rather than integrating over 
$K^1\in \mathbb{R}$ with measure $dK^1/(2\pi)$. Also the $Q^\perp$
contribution is now non-trivial but cancels in the Feynman propagator. 
That is, all expressions 
remain the same except that we must replace $\Delta_+,\Delta_F$ by
\ba \label{6.17i}
\Delta_+(X-Y) &=& \sum_{n\in \mathbb{Z}}\;
\frac{1}{2\omega(n)}\; e^{-i\; K\cdot(X-Y)},\;\omega(n)=2\pi |n|,\;K_1=2\pi n
\nonumber\\
\Delta_F(X-Y) &=& -i\;\int\;\frac{dK^0}{2\pi}\;\sum_{n\in \mathbb{Z}}\;
\frac{e^{-i\; K\cdot(X-Y)}}{-K\cdot K-i\epsilon},\;
\;K_1=2\pi n
\ea
In the so-called ``naive'' discretisation one writes 
\be \label{6.18}
(\partial_M f_M)(m):=\frac{M}{2}[f_M(m+1)-f_M(m-1)]
\ee
for $f_M\in L_M$ the Hilbert space of square symmable 
sequences on the lattice. Using the Fourier functions 
$f_M(m)=e^M_k(m)=e_k(x^M_m)$ with   
$|k|<\pi M$ for $X=\mathbb{R}$ and $f_M(m)=e^M_n(m)=e_{2\pi n}(x^M_m)$ 
with $|n|\le\frac{M-1}{2}$ and $x^M_m=\frac{m}{M}$ with $m\in \mathbb{Z}$ or 
$m\in \mathbb{Z}_M$ respectively   
we find the eigenvalues $\lambda_M(k)$ given by
$i\;M\sin(\frac{k}{M})$ and $iM\sin(\frac{2\pi n}{M})$ 
respectively. These vanish in the allowed domain of $k$ and $n$ respectively 
at $k=0,\; k=\pm \pi M$ and $n=0, n=\frac{M}{2}$ if $M$ is even, otherwise 
only at $n=0$ with corresponding doubler pole in the propagator 
when $K^0=0$. We see that there are no doublers in the compact case 
for lattices with odd numbers of points even with respect to the naive 
discretisation of the discrete derivative. Still, even in the compact case,
and for odd $M$ the eigenvalue $i\;M\sin(\pi\frac{M-1}{M})=-i\;M\;\sin(\pi/M)$
for $n=\frac{M-1}{2}$ approaches $-i\pi$ for large $M$ while most other 
eigenvalues are large of order $M$ and thus $n=\pm(M-1)/2$ can be considered 
as an ``almost'' doubler mode.

We now show that the spectrum of $\partial_M$ is {\it doubler free} if 
we do not pick the naive discetisation but rather the {\it natural 
discretisation} provided by the maps $I_M,\;I_M^\dagger$ in terms of which 
the renormalisation flow is defined. This discretisation is defined by
\be \label{6.19}
\partial_M:=I_M^\dagger\; \partial\; I_M
\ee
for both $X=\mathbb{R}$ and $X=[0,1)$
and is well defined whenever the MRA functions $\chi^M_m$ are at least $C^1$.
Note that with this definition $\partial_M$ is automatically anti-symmetric 
since $\partial$ is.
In fact, for the Haar flow which is not $C^1$ we formally 
find
\be \label{6.20}
\partial_M f_M(m)=
M\;\sum_{\tilde{m}}\;<\chi^M_m,[\chi^M_{\tilde{m}}]'>_L\; f_M(m)=
-M\;\sum_{\tilde{m}}\;<[\chi^M_m]',\chi^M_{\tilde{m}}>_L\; f_M(m)
=\frac{M}{2}(f_M(m+1)-f_M(m-1)
\ee
i.e. precisely the naive derivative where we have formally integrated by 
parts in between and used that $\chi^M_m$ is of compact support for 
$X=\mathbb{R}$ and periodic for $X=[0,1)$ respectively. Thus the Haar 
flow results in the naive discretisation which yields the doubler troubled
spectrum.  

Note that the map $I_M:\; L_M\to L$ has range in $V_M$ and in fact 
$I_M^\dagger:\; L\to L_M$ restricts to the inverse as $I_M^\dagger I_M=1_{LM}$
i.e. $L_M, V_M$ are in bijection. Thus, if in fact $\partial$ preserves 
$V_M$ then the spectrum 
of $\partial_M$ will simply coincide with that of $\partial$ except that 
$k$ will be restricted from $\mathbb{R}$ to $[-\pi M,\pi M]$ and $n$
from $\mathbb{Z}$ to $\mathbb{Z}_M$. This is precisely what happens for 
both the Shannon and Dirichlet kernel as we will now confirm.   

For the Shannon kernel in the case $X=\mathbb{R}$ we compute
\ba \label{6.21}
(\partial_M \; f_M)(m) &=& 
M\;
\sum_{\tilde{m}\in \mathbb{Z}}\; f_M(\tilde{m})\;
<\chi^M_m,\partial\;\chi^M_{\tilde{m}}>_L
(\partial_M \; f_M)(m) 
\nonumber\\
&=& 
M\;
\sum_{\tilde{m}\in \mathbb{Z}}\; f_M(\tilde{m})\;
\int_{\-\pi M}^{\pi M}\;\frac{dk}{2\pi}\; (ik)\;
<\chi^M_m,e_k>_L\; <e_k,\chi^M_{\tilde{m}}>_L
\nonumber\\
&=& 
M\;
\sum_{\tilde{m}\in \mathbb{Z}}\; f_M(\tilde{m})\;
\int_{\-\pi M}^{\pi M}\;\frac{dk}{2\pi}\; (ik)\;
e_{k}(x^M_m-x^M_{\tilde{m}})
\nonumber\\
&=& 
\sum_{\tilde{m}\in \mathbb{Z}}\; f_M(\tilde{m})\;
[\partial_x\; \chi^M_{\tilde{m}}(x)]_{x=x^M_m}
\nonumber\\
&=& 
\sum_{\tilde{m}\in \mathbb{Z}}\; f_M(\tilde{m})\;
[\frac{y\cos(M\pi y)-(M\pi)^{-1}\sin(\pi M y)}{y^2}]_{y=x^M_m-x^M_{\tilde{m}}}
\ea
which displays the non-local nature of the discrete derivative as all points 
$\tilde{m}\in \mathbb{Z}$ contribute. However, (\ref{6.21}) vanishes at 
$m=\tilde{m}$ and takes the maximal value $\mp M$ at $m-\tilde{m}=\pm 1$ 
which shows that it approximates the naive derivative in the vicinity of $m$.
On the other hand, for $f_M=e^M_k$ we find 
the exact eigenfunctions 
\be \label{6.22}
(\partial_M \; e^M_k)(m) = 
M\;
\int_{\-\pi M}^{\pi M}\;\frac{dq}{2\pi}\; (iq)\;
e_{q}(x^M_m)
\sum_{\tilde{m}\in \mathbb{Z}}\; e_{k-q}(x^M{\tilde{m}})\;
=ik\; e^M_k(m)
\ee
with manifestly doubler free spectrum.

For the Dirichlet kernel in the case $X=[0,1)$ the computations are 
completely analogous
\ba \label{6.23}
(\partial_M \; f_M)(m) &=& 
M\;
\sum_{\tilde{m}\in \mathbb{Z}_M}\; f_M(\tilde{m})\;
<\chi^M_m,\partial\;\chi^M_{\tilde{m}}>_L
(\partial_M \; f_M)(m) 
\nonumber\\
&=& 
M\;
\sum_{\tilde{m}\in \mathbb{Z}_M}\; f_M(\tilde{m})\;
\sum_{|n|\le \frac{M-1}{2}}\;(2\pi\;i n)\;
<\chi^M_m,e_{2\pi n}>_L\; <e_{2\pi n},\chi^M_{\tilde{m}}>_L
\nonumber\\
&=& 
M\;
\sum_{\tilde{m}\in \mathbb{Z}_M}\; f_M(\tilde{m})\;
\sum_{|n|\le \frac{M-1}{2}}\;(2\pi\;i n)\;
e_{2\pi \;n}(x^M_m-x^M_{\tilde{m}})
\nonumber\\
&=& 
\sum_{\tilde{m}\in \mathbb{Z}_M}\; f_M(\tilde{m})\;
[\partial_x\; \chi^M_{\tilde{m}}(x)]_{x=x^M_m}
\nonumber\\
&=& 
\sum_{\tilde{m}\in \mathbb{Z}_M}\; f_M(\tilde{m})\;\pi\;
[\frac{\sin(\pi y)\;\cos(M\pi y)-M^{-1}\sin(\pi M y)
\cos(\pi y)}{\sin^2(\pi y)}]_{y=x^M_m-x^M_{\tilde{m}}}
\ea
which displays the non-local nature of the discrete derivative as all points 
$\tilde{m}\in \mathbb{Z}_M$ contribute. However, (\ref{6.23}) vanishes at 
$m=\tilde{m}$ and takes the maximal value $\mp M$ at $m-\tilde{m}=\pm 1$ 
which shows that approximates the naive derivative in the vicinity of $m$.
On the other hand, for $f_M=e^M_n$ we find 
the exact eigenfunctions 
\be \label{6.24}
(\partial_M \; e^M_n)(m) = 
M\;
\sum_{\tilde{m}\in \mathbb{Z}_M}\; e^M_nM(\tilde{m})\;
\sum_{|\tilde{n}|\le \frac{M-1}{2}}\;(2\pi\;i \tilde{n})\;
e^M_{\tilde{n}}(m-\tilde{m})
=2\pi\;i\; n \; e^M_n(m)
\ee
with manifestly doubler free spectrum.\\
\\
We now study the Shannon or Dirichlet flow of the (non-)compact theory.
We start with some initial discretisation $\partial^{(0)}_M,\;
\omega_M^{(0)}=\sqrt{-[\partial^{(0)}_M]^2},\;Q^{(0)}_{M,s}=
\frac{1}{2}[1_{L_M}+i\; s\; \frac{\partial^{(0)}_M}{\omega^{(0)}_M}]$ 
which determines the annihilators in analogy to (\ref{6.14a})
\be \label{6.34}
A^{(0)}_{M,1,+}:=(Q^{(0)}_{M,+} \xi_{M,1})^\ast,\; 
A^{(0)}_{M,1,-}:=Q^{(0)}_{M,-} \xi_{M,1},
A^{(0)}_{M,2,-}:=(Q^{(0)}_{M,-} \xi_{M,2})^\ast,\; 
A^{(0)}_{M,2,+}:=Q^{(0)}_{M,+} \xi_{M,2}
\ee 
the vacuum $\Omega^{(0)}_M$, the Fock space ${\cal H}^{(0)}_M$ 
and the initial Hamiltonian family
\be \label{6.35}
H^{(0)}_M=\sum_{m\in \mathbb{Z}}\; \sum_{B,\sigma}\;
[A^{(0)}_{M,B,\sigma}]^\ast\;\omega^{(0)}_M\; A^{(0)}_{M,B,\sigma}
\ee
and similar for the compact case with the restriction $m\in \mathbb{Z}_M$.

We can encode the flow (\ref{6.32}), (\ref{6.33}) into a single quantity 
$\partial^{(n)}_M$ in terms of which we define analogously
$\omega_M^{(n)}=\sqrt{-[\partial^{(n)}_M]^2},\;Q^{(n)}_{M,s}=
\frac{1}{2}[1_{L_M}+i\; s\; \frac{\partial^{(n)}_M}{\omega^{(n)}_M}]$ 
as well as
\be \label{6.36}
A^{(n)}_{M,1,+}:=(Q^{(n)}_{M,+} \xi_{M,1})^\ast,\; 
A^{(n)}_{M,1,-}:=Q^{(n)}_{M,-} \xi_{M,1},
A^{(n)}_{M,2,-}:=(Q^{(n)}_{M,-} \xi_{M,2})^\ast,\; 
A^{(n)}_{M,2,+}:=Q^{(n)}_{M,+} \xi_{M,2}
\ee 
and the initial Hamiltonian family
\be \label{6.37}
H^{(n)}_M=\sum_{m\in \mathbb{Z}}\; \sum_{B,\sigma}\;
[A^{(n)}_{M,B,\sigma}]^\ast\;\omega^{(n)}_M\; A^{(n)}_{M,B,\sigma}
\ee
and again for the compact case we just restrict to $m\in \mathbb{Z}_M$.
   
To see that this indeed possible we note that in the corresponding Fock
spaces it is sufficient to check isometry on vectors of the form 
\ba \label{6.38}
&& \Psi^{(n)}_{M'}(I_{MM'}\;F_{M,1},..,I_{MM'} \;F_{M,N}):=
A^{(n)}_{M'}(I_{MM'} F_{M,1})^\ast  .. A^{(n)}_{M'}(I_{MM'} 
F_{M,N})^\ast\Omega^{(n)}_{M'},\;A^{(n)}_M(F_M)
\nonumber\\
&:=& \sum_{B,\sigma}
<F_{M,B,\sigma},A^{(n)}_{M,B,\sigma}>_{L_M}
\ea
These give the inner products
\ba \label{6.39}
&&<\Psi^{(n)}_{M'}(I_{MM'}\; F_{M,1},..,I_{MM'}\;F_{M,N}),\;
\Psi^{(n)}_{M'}(I_{MM'}\; G_{M,1},..,I_{MM'}
G_{M,\tilde{N}})>_{{\cal H}^{(n)}_{M'}}
\nonumber\\
&=& \delta_{N,\tilde{N}}
\det([<Q^{(n)}_{M'}\; I_{MM'} F_{M,k},\;
Q^{(n)}_{M'}\; I_{MM'} G_{M,l}>_{L_{M'}^4}]_{k,l=1}^N)
\ea
where 
\ba \label{6.40}
&& <Q^{(n)}_{M'}\; I_{MM'} F_M,\;
Q^{(n)}_{M'}\; I_{MM'} G_M>_{L_{M'}^4}
=\sum_{B,\sigma}\;
<I_{MM'} F_{M,B,\sigma},\;Q^{(n)}_{M'\sigma} I_{MM'} G_{M,B,\sigma}>_{L_{M'}}
\nonumber\\
&=& \sum_{B,\sigma}\;
<F_{M,B,\sigma},\;[I_{MM'}^\dagger\; Q^{(n)}_{M',\sigma} I_{MM'}] 
G_{M,B,\sigma}>_{L_M}
\ea
We used that, whatever $\partial^{(n)}_M$ is, the corresponding operators
$Q^{(n)}_{M,s}\;Q^{(n)}_{M,s'}=\delta_{s,s'}\; Q^{(n)}_{M,s}$
are orthogonal projections and that the $B=1,2$ species anti-commute. 
Comparing with 
\be \label{6.41} 
<\Psi^{(n+1)}_M(F_{M,1},..,F_{M,N}),\;
\Psi^{(n+1)}_M(G_{M,1},..,
G_{M,\tilde{N}})>_{{\cal H}^{(n+1)}_M}
\ee
we obtain isometry iff
\be \label{6.42}
Q^{(n+1)}_{M,\sigma} 
=I_{MM'}^\dagger\; Q^{(n)}_{M',\sigma}\; I_{MM'}
\ee
Similarly, since 
\be \label{6.43}
[H^{(n)}_{M'}\; [A^{(n)}_{M'}(I_{MM'} F_M)]^\ast]=
-[A^{(n)}_{M'}(\omega^{(n)}_{M'}\; Q^{(n)}_{M'} I_{MM'} F_M)]^\ast
\ee
we get match between the matrix elements of Hamiltonians iff
\be \label{6.44}
\omega^{(n+1)}_M
=I_{MM'}^\dagger\; \omega^{(n)}_{M'}I_{MM'}
\ee
where we used that by construction $[\omega^{(n)}_M,Q^{(n)}_{M,s}]=0$. 

We now ask under what conditions on the coarse graining kernel 
$I_M$ both (\ref{6.42}) and (\ref{6.44}) are 
implied by 
\be \label{6.45}
\partial^{(n+1)}_M:=I_{MM'}^\dagger\; \partial^{(n)}_{M'}\;I_{MM'}
\ee
\begin{Theorem} \label{th6.1} ~\\
Suppose that $\partial_M^{(0)}:=I_M^\dagger \;\partial\; I_M$ is the natural
discrete derivative w.r.t. a coarse graining kernel $I_M:\;L_M\to L$ and 
such that 
$[\partial,I_M\;I_M^\dagger]=0$. Then (\ref{6.45}) implies  
both (\ref{6.42}) and (\ref{6.44}).
\end{Theorem}
Proof:\\
By (\ref{6.45}) we have
\be \label{6.46}
\partial^{(1)}_M= [I_{MM'}]^\dagger\;\partial^{(0)}_{M'}\;I_{MM'}
=I_M^\dagger \;\partial\; I_M=\partial^{(0)}_M
\ee
since by construction $I_M=I_{M'}\; I_{MM'}$. Thus 
by iteration $\partial^{(n)}_M=\partial^{(0)}_M=\partial_M$ is already 
fixed pointed, no matter what the coarse graining maps $I_M$ are as
long as they descend from an MRA. 

It follows 
\be \label{6.47}
\partial_M^N=I_M^\dagger\; (\partial\; [I_M \; I_M^\dagger]^{N-1} \;
\partial\; I_M
\ee
While $I_M^\dagger I_M=1_{L_M}$ by isometry, $p_M:=I_M \;I_M^\dagger$ is a 
projection in $L$ (onto the subspace $V_M$ of the MRA). Thus, if 
$[\partial,p_M]=0$ we find $\partial_M^N=I_M^\dagger \partial^N I_M$.
The claim then follows from the spectral theorem (functional calculus).\\
$\Box$\\
\\
To see that both the Shannon and Dirichlet kernel satisfy the assumtion
of the theorem it suffices to remark that they only depend on the difference
$x-y$, i.e. they are translation invariant. Explicitly, since the 
$\chi^M_m$ with $m\in \mathbb{Z}$ and $m\in \mathbb{Z}_M$ respectively 
are an ONB of $V_M$ just as are the $e_k,\; |k|\le \pi M$ and 
$e_{2\pi n},\;|n|\le \frac{M-1}{2}$ respectively  
\be \label{6.48}
(p_M f)(x) = \sum_m \chi^M_m(x)\; <\chi^M_m,f> 
=\left\{ \begin{array}{cc}
\int_X\; dy\; [\int_{-\pi M}^{\pi M}\;\frac{dk}{2\pi}\; e_k(x-y)]\; f(y)
& X=\mathbb{R}\\
\int_X\; dy\; [\sum_{|n|\le \frac{M-1}{2}}\; e_{2\pi n}(x-y)]\; f(y)
& X=[0,1)
\end{array}
\right.
\ee
and integration by parts does not lead to boundary terms due to the 
support properties of $f$ or by periodicity respectively.\\
\\
It follows that using the natural discretisation the theory is already
at its fixed point and the fixed point family member at resolution $M$
coincides with the the continuum theory blocked from the continuum 
to resolution $M$, that is, by simply dropping the superscript $^{(n)}$ we 
have 
\be \label{6.50}
J_M\Omega_M=\Omega,\;
J_M\;A_M(F_{M,1})^\ast..A_M(F_{M,N})^\ast\;\Omega_M,\;
=A(I_M\; F_{M,1})^\ast..A(I_M F_{M,N})^\ast\;\Omega,\;
H_M=J_M^\dagger\; H\; J_M
\ee  
Remark:\\
Thus translation invariance of the Shannon and Dirichlet kernel respectively is,
besides smoothness, another important difference with the Haar kernel
\cite{23}
\be \label{6.49}
\sum_m\; \chi^M_m(x)\chi^M_m(y)= 
=\sum_m\; 
\chi_{[\frac{m}{M},\frac{m+1}{M})}(x)
\chi_{[\frac{m}{M},\frac{m+1}{M})}(y)
\ee
which is not translation invariant. Therefore in this case 
the flows of $\omega_M$ or $\omega_M^{-1}$ are not simply related by 
$\omega_M = I_M^\dagger \omega I_M,\; 
\omega_M^{-1} = I_M^\dagger \omega^{-1} I_M$ and thus one must define 
$\omega_M$ as the inverse of the covariance $\omega_M^{-1}$. As $M\to\infty$
this difference disappears but at finite $M$ it is present and makes the 
study of the flow with respect to a non-translation invariant kernel much 
more and unnecessarily involved.

\section{Conclusion and Outlook}
\label{s5}

In this paper we have extended the definition of Hamiltonian 
renormalisation in the sense of \cite{9} from the bosonic to the fermionic 
case. The definition given in \cite{9} in fact covers both 
cases but the practical implementation for bosons was in terms of measures
\cite{15} which cannot be used for fermions. We have tested the scheme 
for massless 2d chiral fermion theories, the extension to the massive and 
higher dimensional case being immediate, just requiring the higher dimensional
Clifford algebra. In particular we showed that using the smooth local
Shannon-Dirichlet kernel for renormalisation and discretisation results 
in simple flow, an easy computable fixed point theory which coincides with 
the known continuum theory and has manifestly 
doubler free spectrum even at finite 
resolution due to the inherent non-locality with respect to the chosen
finite resolution microscopes based on those kernels.   

An immediate extension of the current paper that suggests itself is 
to apply the current framework to the known solvable 2d interacting fermion 
theories \cite{25}.

\end{document}